\documentclass[runningheads]{llncs}
\usepackage[T1]{fontenc}
\usepackage{graphicx}
\usepackage{hyperref}
\usepackage{booktabs}
\begin{document}
\title{IntellectSeeker: A Personalized Literature Management System with the Probabilistic Model and Large Language Model}

\author{Weizhen Bian\inst{1,3*}\and
Siyan Liu\inst{3}\thanks{These authors contributed equally.} \and
Yubo Zhou\inst{2*}\and
Dezhi Chen\inst{2} \and 
Yijie Liao\inst{2} \and \\
Zhenzhen Fan\inst{3} \and 
Aobo Wang\inst{3}}

\institute{
Hong Kong University of Science and Technology \and
Hong Kong Baptist University \and
National University of Singapore
}
\authorrunning{W. Bian et al.}

\maketitle 

\begin{abstract}
Faced with the burgeoning volume of academic literature, researchers often need help with uncertain article quality and mismatches in term searches using traditional academic engines. We introduce IntellectSeeker, an innovative and personalized intelligent academic literature management platform to address these challenges. This platform integrates a Large Language Model (LLM)--based semantic enhancement bot with a sophisticated probability model to personalize and streamline literature searches. We adopted the GPT-3.5-turbo model to transform everyday language into professional academic terms across various scenarios using multiple rounds of few-shot learning. This adaptation mainly benefits academic newcomers, effectively bridging the gap between general inquiries and academic terminology. The probabilistic model intelligently filters academic articles to align closely with the specific interests of users, which are derived from explicit needs and behavioral patterns. Moreover, IntellectSeeker incorporates an advanced recommendation system and text compression tools. These features enable intelligent article recommendations based on user interactions and present search results through concise one-line summaries and innovative word cloud visualizations, significantly enhancing research efficiency and user experience. IntellectSeeker offers academic researchers a highly customizable literature management solution with exceptional search precision and matching capabilities. The code can be found here: https://github.com/LuckyBian/ISY5001

\keywords{knowledge management  \and knowledge engineering \and large language model(LLM) \and few-short learning \and probabilistic model \and recommendation system \and chat robot.}

\end{abstract}

\section{Introduction}

With the overwhelming growth of research documents in digital libraries, finding relevant material has become increasingly time-consuming. This surge in academic content demands the creation of intelligent systems to streamline the literature search process\cite{sharma2023anatomization}. In response to this challenge, the IntellectSeeker platform has emerged as a pioneering tool designed to revolutionize how researchers navigate the expansive realm of scholarly articles, especially for those newly initiated into academia.\\

Central to IntellectSeeker's innovation is a Large Language Model-based semantic enhancement bot. This cutting-edge feature is fine-tuned to transform everyday language into academic vocabulary, bridging the gap between general queries and scholarly terminology. IntellectSeeker also introduces a sophisticated probabilistic model\cite{porba} for data crawling, which effectively balances user-defined criteria with implicit user preferences derived from their activities on the platform. This model dynamically tailors the database to match academic articles closely with the user's evolving research interests. By analyzing explicit user requirements and subtle behavioral patterns, the system selects and retrieves the most pertinent academic articles through probabilistic methods, thus personalizing the research experience.\\

With these features, IntellectSeeker further enhances the user experience with complementary tools. The platform features an advanced recommendation system that intelligently recommends articles, deriving insights from user interactions such as likes, bookmarks, and reading history. Additionally, the platform includes text compression tools that generate one-line summaries and innovative word cloud visualizations to enhance search result analysis. In the following chapters, we will introduce the system in detail.

\section{Related Works}
Literature management systems have evolved significantly, transitioning from traditional document management tools to more innovative and personalized advanced systems. Traditional academic search engines like ScienceDirect, ResearchGate, and Google Scholar offer basic recommendations but often must accurately meet some researchers' needs. Even if these systems work correctly, their suggestions are only sometimes relevant or valuable for the user. It is also essential to be cautious with these platforms, especially for citations and metrics, as they can be manipulated and sometimes have issues with the quality of their indexing\cite{halevi2017suitability}. We have noticed a shortage of highly customizable academic literature management systems that can construct targeted literature databases based on individual research interests.\\

In the domain of personalized literature management systems, particularly in recommending academic papers, various methods are employed by recommendation systems to address cold-start scenarios. Standard methods include Content-Based Filtering (CBF), Collaborative Filtering (CF), Link-Based and co-occurrence-based Approaches, and various hybrid methods. In the field of academic paper recommendation, various methodologies are utilized. Content-based filtering typically employs the Bag-of-Words model\cite{lee2013personalized}, where term weights are represented as binary values or term frequencies\cite{sugiyama2010scholarly}, and the Term Frequency-Inverse Document Frequency (TF-IDF) method is used\cite{chaitanya2017research}. Additionally, it has been found that the distributed representation method 'doc2vec' outperforms other text representation techniques\cite{zhang2022citation}. The method primarily relies on a rating database for collaborative filtering approaches, recommending papers that similar users have highly rated. However, rating an article requires thorough reading and understanding, a time-consuming process often leading to a limited number of ratings due to slow user response\cite{yang2009cares}. The heavy reliance on user participation leads to the cold-start problem, which can be split into user and item cold-starts\cite{sharma2017collaborative}. To mitigate reliance on explicit ratings, implicit ratings can be inferred from user actions such as downloading\cite{guo2017pagerank}, adding papers to a collection\cite{alotaibi2013trust}, and commenting\cite{guo2017pagerank}. In link-based methods, a research document can be represented by its title, unique paper ID, and other information like the author's name\cite{zhou2008learning}, type\cite{safa2018publication}, and publisher\cite{ma2019personalized}. These methods enable the system to link documents comprehensively, thereby enhancing the information available to users.\\ 

Large Language Models (LLMs) are breaking new ground by translating everyday vocabulary into academic terms, a task for which literature management systems currently need more established precedents. Common challenges include inaccuracies in literature search results, often stemming from unfamiliarity with specialized academic terminology in new fields. To address this, LLMs in our model undergo extensive training on diverse text datasets, followed by fine-tuning through meta-training to align with human preferences. This foundational training equips the models with a robust linguistic understanding, significantly enhancing their performance in academic applications\cite{kalyan2021ammus}.\\

The advent of LLMs began with the introduction of GPT-3\cite{brown2020language}, successfully leading to the development of various other LLMs. These include models like PaLM\cite{chowdhery2023palm}, PaLM2\cite{anil2023palm}, LaMDAMDA\cite{thoppilan2022lamda}, Megatron–Turing NLG\cite{smith2022using}, and LLaMA\cite{touvron2023llama}. The field of LLMs has seen exponential growth, especially following the recent release of advanced models by OpenAI, such as ChatGPT and GPT-4\cite{openai2023gpt}, which have significantly increased their popularity. The successful development of LLMs has also led to the creation of numerous models for specialized domain applications, such as BloombergGPT\cite{wu2023bloomberggpt} developed by Bloomberg in the finance domain, CodeGen2\cite{nijkamp2023codegen2} in the programming domain, MedPaLM\cite{singhal2023large} in medical domain, Goat\cite{liu2023goat} in education domain. \\

In question-answering tasks, Researchers in the NLP field have investigated OpenAI-based General Large Language Models for question-answering applications across different domains. These include Visconde\cite{pereira2023visconde} in literature understanding, social media\cite{ye2023comprehensive}, zero-shot and few-shot in news sector\cite{srivastava2022towards}, etc., These studies all employed zero-shot or one-shot learning methods and demonstrated promising results. However, there is no need for a specialized fine-tuned GPT tool specifically for addressing academic term matching issues encountered in academic searches, nor is there an implementation of its lightweight training method and application.\\

Aiming at existing research problems and the current industry situation, this paper proposes a personalized literature management system with a probabilistic and large language model to solve them. The main contributions of this paper include:
\begin{enumerate}
    \item We provide a personalized literature management platform that facilitates targeted literature curation and ensures the selection of high-quality academic papers. This addresses the need for a more user-centric approach in literature management, catering to individual research interests and preferences.
    \item We added the Probabilistic Model to scrap data. The Probabilistic Model will disassemble the scraped articles, match the disassembled features with the user's active input needs and the collected user activities, and then calculate the probability to decide whether the article is collected. Therefore, as the use of IntellectSeeker increases, the database will gradually be personalized.
    \item We have developed an intelligent chatbot system using a fine-tuned GPT model. This system is designed to assist in accurately translating everyday language into specific academic terminology in a professional format, thereby enhancing the precision of search results in academic contexts, especially for users unfamiliar with certain fields.
\end{enumerate}

\section{Proposed System}
\subsection{System Overview}
\begin{figure}
\includegraphics[width=\textwidth]{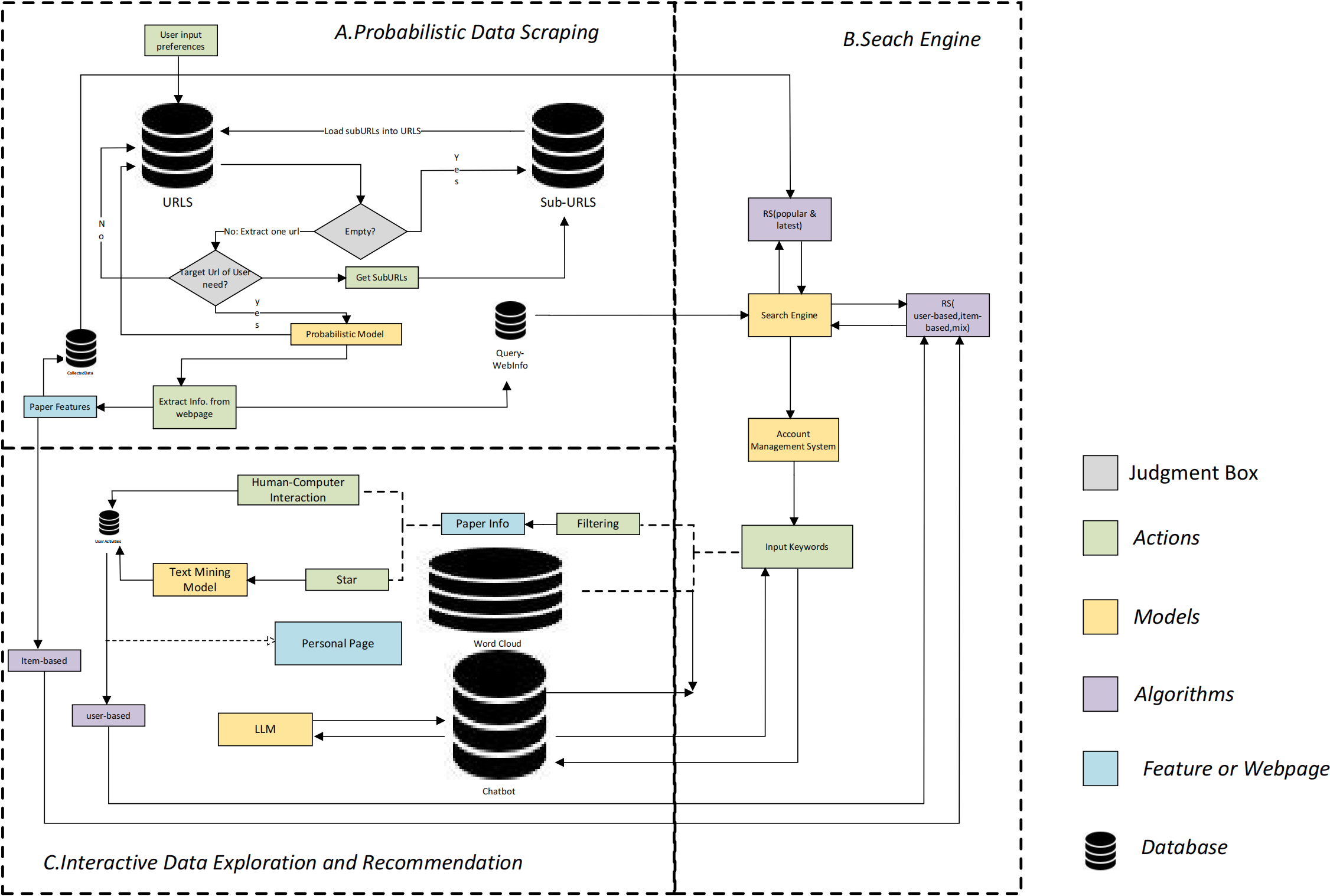}
\caption{Overview of IntellectSeeker: IntellectSeeker features three core components: Probabilistic Data Scraping, Enhanced Search Engine, and Interactive Data Exploration and Personalized Recommendation. It uses probabilistic algorithms to filter web data that are closely aligned with user preferences. The search engine dynamically adjusts search results based on user queries and preferences, incorporating popular and personalized suggestions. The interactive module uses a text mining model and an LLM-based chatbot to deliver customized recommendations and visual summaries like word clouds, enhancing the user experience through tailored content.} 
\label{fig1}
\end{figure}

Fig. \ref{fig1} showcases the IntellectSeeker platform, a sophisticated academic tool that blends data scraping, an enhanced search engine, and an interactive interface to revolutionize academic research. A probabilistic model, central to its data scraping mechanism\cite{porba}, is an essential aspect of Component A. This model adeptly selects scholarly articles to get a personalized database, intelligently aligning the data-gathering process with user-defined preferences. The academic articles that the user needs are accurately selected by considering the user's various preferences and features. Component B, the search engine, excels in query refinement and incorporates an advanced ranking system. This system judiciously assesses factors such as the timeliness of documents, user preferences, and historical user interactions, thereby delivering a well-tailored and pertinent set of documents\cite{search}. Lastly, component C, which focuses on interactive data exploration and recommendation, harnesses the power of large language models (LLMs) to refine and enhance the search experience\cite{zhong2023can}. This subsystem is adept at interpreting nuanced user queries and translating them into academic lexicon, thus optimizing the search term entries. By analyzing user interactions such as likes and bookmarks, the IntellectSeeker platform contributes to developing a highly personalized user interface. This interface visualizes the search result by crafting detailed word clouds that distill and represent the core themes of the explored literature\cite{cui2010context}. \\

\subsection{Probabilistic Data Scraping Model}

\begin{figure}
\includegraphics[width=\textwidth]{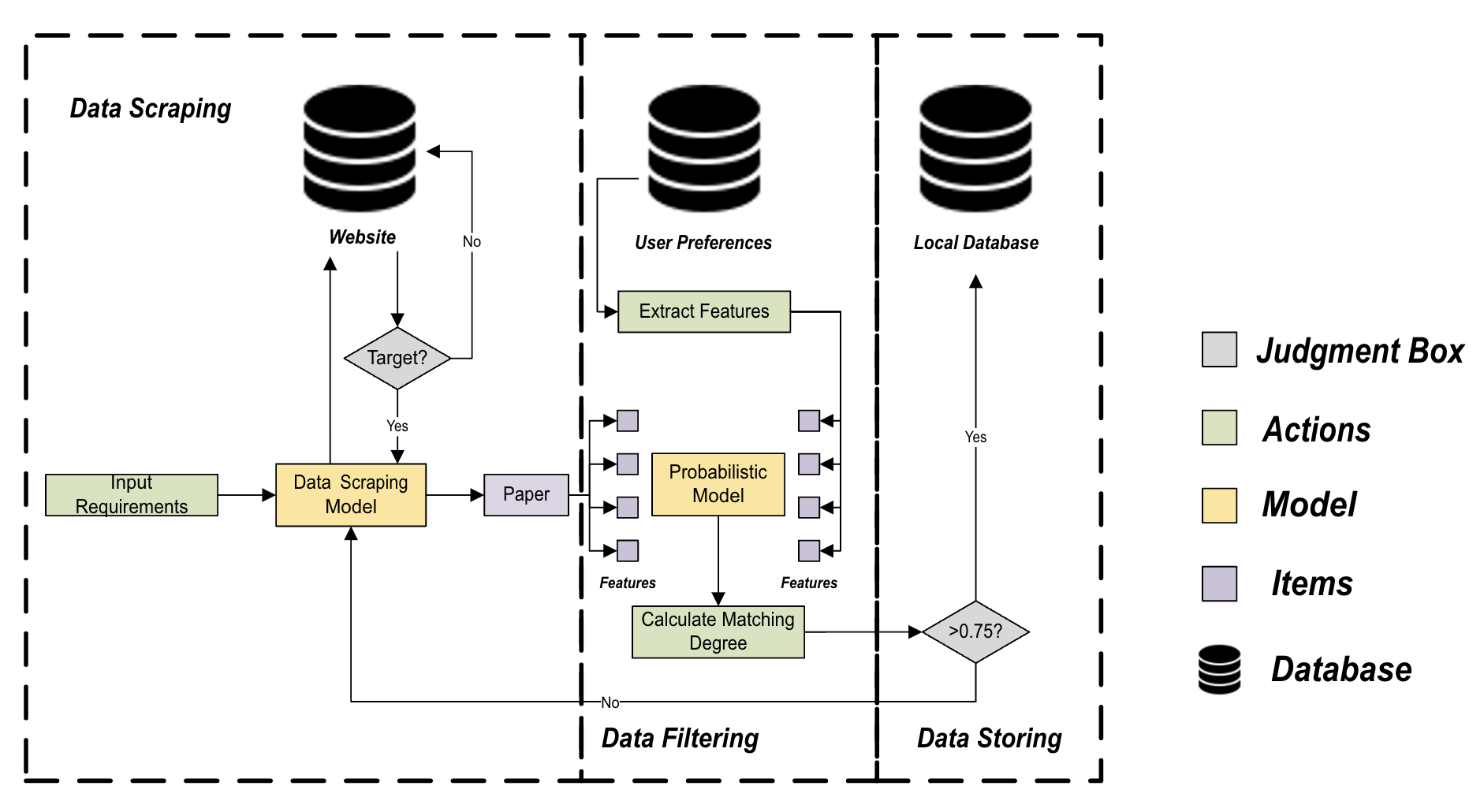}
\caption{IntellectSeeker's data crawling process: This diagram illustrates IntellectSeeker's data crawling process, divided into three key stages. First, user input guides the system in extracting relevant data from the target website using a probabilistic model. Subsequently, the extracted data is evaluated against predefined thresholds to determine its relevance. Finally, it is stored through pointers to ensure storage space conservation.} \label{data}
\end{figure}

In academic research, personalization and precision in data retrieval are crucial\cite{personal}. Therefore, we designed a probabilistic data scraping model (shown in Figure \ref{data}) to create a personalized database\cite{datam}. In the IntellectSeeker data crawling model, in addition to the traditional academic search platform's filtering function for factors such as document year and journal\cite{gusenbauer2020academic}, it also adds factors such as user preferences to locate user needs accurately. As shown in figure \ref{data}, IntellectSeeker allows users to impose restrictions on the data scraping process during the data scraping stage, allowing the exclusion of documents in specific journals or conferences, thereby filtering out low-quality academic publications. Additionally, users can manually fine-tune data scraping, and the system dynamically adjusts its recommendations accordingly. A unique feature of IntellectSeeker in this process is using a hashing algorithm\cite{ha} to assign a unique web identifier (webid) to each article.\\

In the advanced stages of data filtration, the IntellectSeeker platform employs a refined approach, drawing inspiration from decision tree\cite{song2015decision} algorithms. This process entails meticulously matching article features with user-specific characteristics, executed via a probabilistic model. The operational essence of this model is encapsulated in the following formula:

\begin{equation}
    I = w_p \cdot S(K_a, K_u) + w_i \cdot S(K_a, K_i)
    \label{eq:importance_score_decision_tree}
\end{equation}
where \(I\) is the importance score, \(K_a\) represents the article's features, \(K_u\) denotes the user's preference features, \(K_i\) signifies the user's manually inputted requirements, \(w_p\) and \(w_i\) are weights for user preferences and manual inputs respectively, and \(S\) is the similarity function measuring the overlap between sets of keywords. `I' represents the calculated importance score, ranging from 0 to 1.\\

In the above formula, all features K will be converted into codes, then the similarity will be calculated. In addition, users can manually adjust the weights to adjust the impact ratio of input requirements and historical record preferences. In addition, the platform is configured with a predefined threshold of 0.75, and academic articles with importance exceeding this value are directly saved in the database. In addition, a probability value of P=0.05 is set during data collection so that some articles with a threshold lower than 0.75 also have a certain probability of being collected, expanding the range of research literature users can access.

\subsection{Search Engine}
\begin{figure}
\centering
\includegraphics[width=\textwidth]{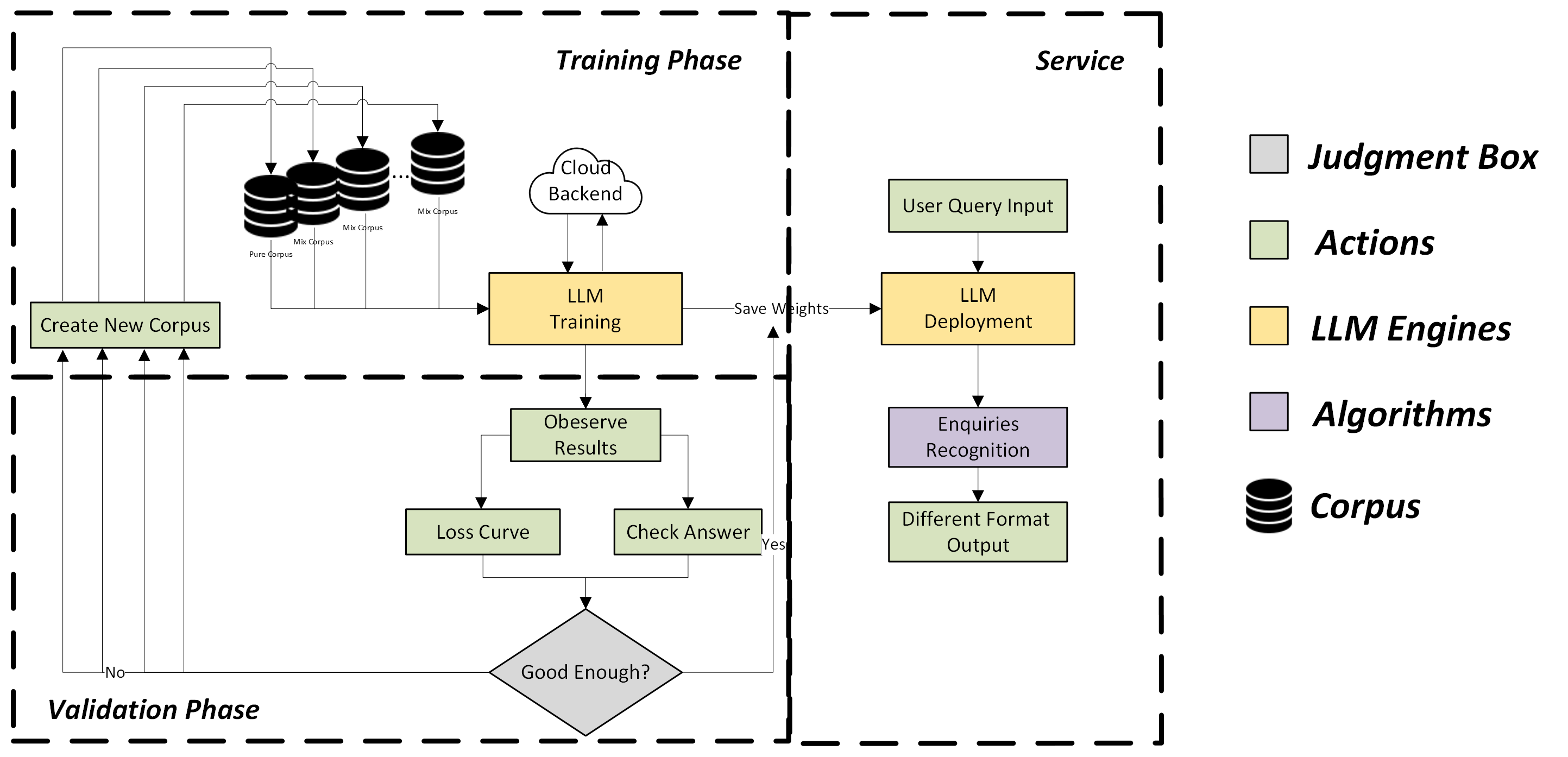}
\caption{Overview of LLM Search Engine: it outlines a Large Language Model's training and deployment workflow. During the 'Training Phase,' a new corpus is created and used to fine-tune a pre-trained large language model. Once validated, the model is deployed to handle user queries, recognizing and processing inquiries into various formats. Key components include the 'Judgment Box' for response evaluation, 'Actions' guiding system responses, 'LLM Engines' for processing, 'Algorithms' for enhanced functionality, and the 'Corpus' as the training dataset.} \label{search}
\end{figure}
Within the IntellectSeeker, the Semantic Enhancement with Large Language Model is a pivotal component crafted to deal with the problems users may encounter with academic language during scholarly searches. This subsystem's architecture and rationale are deeply rooted in understanding the user’s query process and the wide range of terminology used across various academic realms. Researchers new to a field may use general terms due to unfamiliarity with the specialized vocabulary, which often leads to irrelevant results when using standard search engines like Google Scholar. To address this issue, we have deployed a semantically enhanced fine-tuned LLM dedicated to tackling this challenge as shown in Fig. \ref{search}.\\

This approach affords access to an extensive and interactive academic lexicon that is exceptionally professional and knowledge-rich, ensuring that even those researchers new to a field can swiftly acquire relevant academic term replacements, significantly enhancing the accuracy and efficiency of academic searches. The ability of natural language understanding and knowledge of LLM also ensure that it can understand various natural language instructions accurately, catering to different needs. It can transform vague terms into a series of alternatives across various fields or offer exact replacements in specific domains if users ask for them or simply define the domain. Through the LLM, users gain a personalized search experience with consistent, user-friendly, and easy-to-read outputs.\\

We have carried out a series of experiments on different LLMs, including open-source models like Llama2\cite{touvron2023llama} families, Gemma2\cite{team2024gemma}, Mistral-7b\cite{jiang2023mistral} and OpenAI model gpt-3.5-turbo\cite{ouyang2022training}. For the OpenAI model, we experimented with different shot instruction strategies and fine-tuning to enhance the performance. We compared all LLMs and finally utilized a finetuned OpenAI GPT-3.5-turbo model based on a set of metrics that more perfectly matches the needs of this system component, providing more stable and comprehensive outputs, especially in the conversion from daily vocabulary to academic vocabulary across various research disciplines. We further crafted a user-friendly dialogue mode and framework through manual construction, which includes setting the format and style of responses, handling specific user requests such as academic term substitutions within designated fields, and addressing the reliability of term explanations. 

\subsection{Interactive Data Exploration and Recommendation}

\subsubsection{Automatic Summarization}
In the IntellectSeeker platform, a key feature is its sophisticated summarization capability, which employs the SMMRY API\cite{api} to distill voluminous article abstracts into concise statements. This functionality is pivotal for accelerating the research process, particularly for PhD scholars who require rapid filtering and prioritization of relevant literature. Upon the user bookmarking a webpage, the system activates the API, which then scans and truncates the content, eliminating non-essential elements. This process embeds coherent, compressed summaries in saved pages, helping users filter out unwanted documents more quickly.

\subsubsection{Information Visualization}
The IntellectSeeker platform has harnessed the power of visualization to enhance user experience during the search process. Upon initiating a query, the platform comprehensively analyzes the search results, aggregating the core terms identified within these documents\cite{cui2010context}. Subsequently, it extracts the twenty most frequently occurring terms from this collective pool, weaving them into an informative word cloud. This visual representation serves as a snapshot of the prevalent themes and concepts within the search results, allowing users to ascertain the pertinence of the information retrieved quickly. If discrepancies exist or the results do not align with the intended search criteria, the word cloud provides immediate visual feedback. 

\subsubsection{Recommendation}
The IntellectSeeker platform refines the academic research experience with its recommendation system. Drawing from traditional collaborative filtering methods, it analyzes user interactions such as likes, bookmarks, and clicks to recommend similar literature or connect users with similar research interests\cite{zhao2010user}\cite{deshpande2004item}. Additionally, it incorporates trending articles and recent publications to ensure researchers have access to the most current and influential works\cite{steck2011item}. A hybrid algorithm integrates these approaches, weighting each to produce a tailored and dynamic list of literature, thereby streamlining the discovery process for scholars\cite{lucas2013hybrid}.

\section{Implementation}

\subsection{Train Semantic Enhancement Model}
\subsubsection{Data Corpus Construction}
We built an initial corpus with definitions of system roles and question-and-answer pairs. 0.266k rows of sample pairs were made for academic replacements of daily expressions specifying a particular field; the output was in a fixed format, providing academic replacement words and their definitions. In fine-tuning the models, the dataset was split into 9:1 as a training set and validation set. We manually wrote several prompts to construct this dataset, with academic word definitions sourced from academic databases \cite{coxhead2000new}. These manually written prompts were then used to guide ChatGPT in generating more question-and-answer pairs in the same format defined manually. For the common unrelated questions, we allowed ChatGPT to generate them randomly, add variety to the corpus, and improve the model's adaptability to various input types.

\subsubsection{Fine-tuning and Iterative Learning}
Considering the efficiency and cost of the system, we didn't use open-source LLM with supermassive parameters. Table \ref{tab:experimental_results} shows the result of different LLMs, including Llama2 with 7B and 13B, Gemma and Mixtral with the highest MT-bench score. The training methods include fine-tuning, zero-shot, one-shot, and few-shot instructions guide. For those experiments in which the training corpus is not demanded, we use the same validation set as the fine-tuned one to calculate the metrics. For performance comparison, BLEU, ROUGE series and METEOR which are the typical metrics in text conversion like translation are chosen for evaluation.\\

\begin{table}[htbp]
\centering
\caption{Experimental Results of LLMs}
\label{tab:experimental_results}
\begin{tabular}{cccccccc}
\toprule\small
Model& Actions & BLEU & ROUGE-1&ROUGE-2 & ROUGE-L & METEOR\\
\midrule
Llama2-7B     & fine-tune & 0.0    & 0.1423 & 0.0072 & 0.0983 & 0.1423 \\
Llama2-13B    & fine-tune & 0.0919 & 0.2050 & 0.0418 & 0.1430 & 0.3309 \\
Gemma2-2B     & fine-tune & 0.1933 & 0.2821 & 0.0951 & 0.2052 & 0.4720 \\
Mistral-7B    & fine-tune & 0.0989 & 0.2079 & 0.1258 & 0.1759 & 0.4352 \\
Gpt-3.5-turbo & zero-shot & 0.0	   & 0.2125	& 0.0287 & 0.1949 & 0.1311 \\
Gpt-3.5-turbo & one-shot  & 0.1562 & 0.3066	& 0.0605 & 0.2252 & 0.4966 \\
Gpt-3.5-turbo & few-shot  & 0.4445 & 0.6058 & 0.4517 & 0.5554 & 0.7155 \\
\textbf{Gpt-3.5-turbo} & \textbf{fine-tune} &\textbf{0.9269}&	\textbf{0.9413}&	\textbf{0.9160}&	\textbf{0.9355}	&\textbf{0.9553} \\

\bottomrule
\end{tabular}
\end{table}

 The results demonstrate that the fine-tuned GPT-3.5-turbo model outperforms others in all metrics, achieving a BLEU score of 0.9269, ROUGE-1, ROUGE-2, and ROUGE-L scores of 0.9413, 0.9160, and 0.9355 respectively, and a METEOR score of 0.9553. These results indicate that the fine-tuned GPT-3.5-turbo model meets our system requirements more perfectly in terms of stability and comprehensive output and also excels in converting everyday and academic vocabulary. Its larger model parameters and extensive academic vocabulary knowledge across various research fields provide a superior user experience. Therefore, we ultimately selected the fine-tuned GPT-3.5-turbo model as our primary model to ensure our system's best performance and user experience.

\subsection{Improve data Scraping Speed}
In its quest to optimize data scraping efficiency, the IntellectSeeker platform integrates a sophisticated multi-threaded approach, significantly expediting the data retrieval process\cite{edelstein2003framework}. This is adeptly complemented by applying advanced Natural Language Processing (NLP) techniques, which are crucial in refining query mappings\cite{tabassum2020survey}. These techniques involve strategically removing stop words and implementing lemmatization algorithms, effectively streamlining linguistic data processing. This minimizes the storage requirements and markedly enhances the speed and precision of search operations. Furthermore, the platform adopts an innovative pointer-based storage system meticulously engineered to circumvent the pitfalls of redundant data storage\cite{chilimbi2000making}. This system not only preserves the integrity and organization of the stored data but also significantly contributes to the overall efficiency of data management within the platform.

\section{Conclusion and Future Work}
IntellectSeeker is a major step forward in scholarly research tools, providing unparalleled ease and precision when browsing vast literature databases. Its core capabilities include sophisticated probabilistic data crawling mechanisms and large language models to improve search precision. While IntellectSeeker significantly enhances literature management, there is room to enhance the capabilities of LLM to include more comprehensive scholarly services, such as advanced question-answering. Further innovations will be made with IntellectSeeker. Future efforts should focus on a deeper analysis of user interactions to create detailed profiles that lead to more targeted search strategies.

\bibliographystyle{unsrt}
\bibliography{ref}

\end{document}